\documentclass[12pt,a4paper]{article}
\usepackage{amssymb,amsmath,latexsym}
\makeatletter
\usepackage{appendix}

\usepackage[font={small,it}]{caption}
\usepackage[T1]{fontenc}
\usepackage[utf8]{inputenc}
\usepackage{physics}
\usepackage{lmodern}
\usepackage{amsmath}
\usepackage{amssymb}
\usepackage{graphicx}
\usepackage{color}
\usepackage{amsmath}
\newcommand{\be}{\begin{equation}}
\newcommand{\ee}{\end{equation}}
\newcommand{\ba}{\begin{eqnarray}}
\newcommand{\ea}{\end{eqnarray}}

\title{Black Hole Entropy and Boundary Conditions}

\date{}
\begin{document}
\maketitle
\begin{center}
\author{H. Khodabakhshi $^{1,4,6}$, A. Shirzad $^{3,4}$, F. Shojai $^{1,2}$, Robert B. Mann$^{5,6}$
	\\
	\small$^1$Department of Physics, University of Tehran, Tehran, Iran\\
	\small$^2$Foundations of Physics Group, School of Physics, Institute for Research in Fundamental Sciences (IPM), Tehran, Iran\\ \small$^3$Department of Physics, Isfahan University of Technology, Isfahan, Iran\\ \small$^4$School of Particles and Accelerators, Institute for Research in Fundamental Sciences (IPM), Tehran, Iran\\ \small$^5$Perimeter Institute, 31 Caroline St. N., Waterloo, Ontario, N2L 2Y5, Canada\\ \small$^6$Department of Physics and Astronomy, University of Waterloo, Waterloo, Ontario, Canada, N2L 3G1
}
\end{center}
\begin{center}
\hrulefill

\begin{abstract}
It is well-known that in order to make the action well defined, one may employ different kinds of boundary conditions (BCs) accompanied by the appropriate Gibbons-Hawking-York (GHY) terms. In this paper we investigate the role of the selected BC and the corresponding GHY terms on the black hole (BH) entropy. Our result shows, regardless of the kind of BC, the BH entropy in all cases is the same as one obtained under Dirichlet BC from Wald formula or semi-classical approximation method. We considered the Schwarzschild solution for $f(R)$-gravity and general relativity (GR) in standard dimensions as special models.
	
\end{abstract}
\hrulefill
\end{center}

\section{Introduction}

There are two known methods for finding black hole (BH)  entropy for a given solution of the equations of motion in a gravitational theory, i.e. the Wald method \cite{1,2,3,4} and the Euclidean semi-classical approximation \cite{5,6}. The Wald method is based on calculating an integral over the bifurcation two-sphere surface, where the integrand comes from the Lagrangian density \cite{7}. Under certain conditions, such as  Dirichlet boundary conditions (BCs), the Wald method leads to the famous result of Bekenstein-Hawking \cite{8}. To obtain Wald formula it is necessary to make the action well-defined, under certain BCs.   In the Euclidean semi-classical approximation, one calculates the difference between  the Euclidean action for a given solution and an appropriate background metric. In this way one obtains the partition function, the free energy, and the entropy of the solution. As with the Wald method, the action should support a well-defined variational principle in the Euclidean semi-classical approximation method as well \cite{13}. 

It is well-known that the surface integrals generated by  variation of the action in a gravitational theory cannot be omitted simply by imposing BCs. It is necessary to supplement the action by adding appropriate Gibbons-Hawking-York (GHY) terms \cite{6,14,15,16}.  These terms play an important role in calculation of the BH entropy \cite{22,23,24,25}. However, the GHY terms take different forms under different BCs;
Dirichlet BCs are not the only ones required to make the action principal well-defined. One may use some other BCs such as Neumann or mixed BC, accompanied by appropriate GHY terms. 

These issues are particularly pertinent in theories of gravity that are higher order in curvature. The Wald method has been used in deriving  BH entropy in modified gravity models such as $f(R)$-gravity \cite{9,10} and higher curvature gravity under Dirichlet BCs \cite{5,11,12}.   More generally, to make the variational principle well-defined in a higher order Lagrangian, one must first render the Lagrangian degenerate. Some Lagrangians, such as the Einstein-Hilbert Lagrangian, are automatically degenerate (i.e. we can write them as the sum of quadratic parts in the first order derivatives of metric and a total derivative term).  However others, like the $f(R)$-gravity Lagrangian, are not \cite{17}. These latter cases can be made degenerate via the Ostrogradsky approach \cite{15}. 
In the Wald and semi-classical approximation methods we will consider this important point in what follows.

The question then arises as to what happens to BH entropy under different BCs as well as different GHY terms. 
 In this paper we follow the program of Wald and the Euclidean semi-classical approximation to find the BH entropy for asymptotically flat theories. We expect physically that the entropy should be the same under different BCs and our results indeed show that  the  origin of  BH entropy is the same in all cases.
 
Amongst the many higher-curvature theories available, we shall concentrate our attention on $f(R)$-gravity. This is partly because $f(R)$-gravity provides an  important example of the arbitrariness in choice of BCs \cite{13,15}; indeed, in some cases no GHY term is even necessary. More generally,
$f(R)$-gravity is the simplest class of higher-curvature theories of gravity that have been of physical interest in recent years \cite{new9,new3,new4}, particularly in cosmology, where they have been shown to  explain cosmic acceleration without dark energy see\cite{new5,new6,new7,new8}.  Conceptually, they generalize the hypothesis that the Einstein-Hilbert action for the gravitational field is linear in the Ricci scalar $R$, to an action that is some general function $f(R)$ of this quantity.

In section 2 we   give a review of the Wald method and we will obtain the entropy formula for different BCs in higher curvature gravity. As an example we  consider  $f(R)$-gravity and its special case GR.  We show that although the Noether current and Noether charge change under different BCs through the GHY terms,  BH entropy for an asymptotically flat solution turns out to be the same for different BCs.

In section 3 we  consider the semi-classical approximation method for $f(R)$-gravity, where in addition to Dirichlet BCs, the entropy is computed for Schwarzschild metric in flat background for Neumann BC and two types of mixed BCs. Similar to the Wald method we will show, despite the fact that GHY surface terms are different for different BCs, the entropy  turns out to be the same.  We show how to reduce our results to the GR case.

In all of our manipulations, we follow a covariant approach for making the variational principle well-defined. Then we compute the BH entropy by using the ADM coordinate system. However, it is also possible to compute the entropy without going to the ADM formulation.  Latin indices are used for space-time coordinates and  Greek ones are  denote the spatial coordinates. 

\section{Wald Entropy Under Different BCs}

 The Noether charge method employs the Lagrangian density to obtain the entropy of a BH. The final result is  \cite{7}
\begin{align}
\mathrm{S}=-2\pi\int_{\mathcal O} \frac{\partial\mathcal{L}}{\partial R_{abcd}} \hat{\epsilon}_{ab}\hat{\epsilon}_{cd}\bar{\epsilon}, 
\label{wald}
\end{align}
in which $\mathcal{L}$ is the Lagrangian density, $\hat{\epsilon}_{ab}$ is the bi-normal vector on the bifurcation two-sphere $\mathcal{O}$ and $\bar{\epsilon}$ denotes the surface element.  

It appears from (\ref{wald}) that only the Lagrangian density $\mathcal{L}$ is required to obtain $S$; there is no need to make the action well-posed.
In this section we address the question of whether or not   a well-posed action is required to obtain (\ref{wald}). In other words, is it necessary to add appropriate GHY terms before obtaining the BH entropy?  We shall show that 
while  we must render the action well-posed under certain BCs to obtain (\ref{wald}),  changing BCs cannot affect the final result. 
 
Let us consider a higher derivative Lagrangian: $\mathcal{L}=L(\Phi, \partial \Phi, \partial^{2} \Phi,...)$ in which $\Phi$s are dynamical fields, including the metric.  A higher order Lagrangian does not necessarily  have a well-posed variational principle and also a well-defined Hamiltonian structure. Therefore, if the Lagrangian is not degenerate, we must decrease the order of derivatives via the Ostrogradsky approach \cite{17}. In this approach we decrease the order of derivatives by increasing the number of degrees of freedom. We assume the Lagrangian $L$ has been made degenerate so that the action may be written as
\begin{align}
\mathcal{A}=\int_\nu L-\int_{\partial \nu}l.
\label{1}
\end{align}
where the additional surface terms in \eqref{1} are the GHY terms for a certain BCs\footnote{We note that for some Lagrangians there is no need to add any GHY terms and so $l=0$. For example in an asymptotically flat space-time, under Neumann BCs in four dimensions for GR and under a kind of mixed BC in D dimensions for $f(R)$-gravity, we do not need any GHY terms to make the action of these theories well-defined 
\cite{15}.}. Also the boundary $\partial \nu $ may be seen as the union $(-\Sigma_{t_1} ) \bigcup \Sigma_{t_2}\bigcup \Gamma $ where $\Sigma_t$s denote the spacelike hyper-surfaces and $\Gamma$ denotes a timelike hyper-surface.  In general, by varying the Lagrangian $ L $ with respect to the fields $\Phi$, we have
\begin{align}
\delta L = E_{a} \delta \Phi^{a}+ d\theta (\delta \Phi),
\label{2}
\end{align}
in which $\theta$ depends on $\Phi$,$\delta \Phi$ and their first derivatives. Varying the action (\ref{1}) and using above formula we obtain
\begin{align}
\delta \mathcal{A}=\int_{\mathcal{M}}E_{a}\delta \Phi ^a +\int_{\partial \mathcal{M}}(\theta-\delta l).
\label{3}
\end{align}
To obtain a well-defined variational principle, we 
select an appropriate BC for which $\theta=\delta l$ and impose the least action principle, yielding 
the equations of motion $E_{a}=0$.  One can introduce a Noether current 
associated with an infinitesimal diffeomorphism  and obtain  the corresponding Noether charge \cite{7}. An infinitesimal variation of the total Lagrangian $L'=L-dl$ due to an infinitesimal diffeomorphism generated by a vector field $\xi$ yields 
\begin{align}
d(\theta(\delta_{\xi}\Phi)-\xi\cdot L'-\delta_{\xi}l)=-E_{a}\delta_{\xi}\Phi^a,
\label{4}
\end{align}
where we have used Cartan’s magic formula \cite{13}   and the relation
\begin{align}
\delta_{\xi}L=E_a \delta_{\xi} \Phi^a +d \theta (\delta_{\xi} \Phi).
\label{5}
\end{align} 
 Imposing the equation of motion $E_{a}=0$ in (\ref{4}), we can introduce the conserved current
\begin{align}
J'_{\xi}= \theta (\delta_{\xi}\Phi)- \xi\cdot L'-\delta_{\xi}l,
\label{6}
\end{align} 
for which  $dJ'_{\xi}=0$. Locally this implies $J'_{\xi}=dQ'[\xi]$ and the associated conserved Noether charge  can be obtained from this.

To obtain the entropy from this Noether charge we vary Eq.(\ref{6}), impose $E_{a}=0$, and set $\theta=\delta l$,
obtaining 
\begin{align}
\delta J'_{\xi}= \delta \theta (\delta_{\xi}\Phi)- \delta_{\xi} \theta (\delta \Phi)
\label{7}
\end{align} 
using \eqref{2}, where  we assumed $\delta$ commutes with $\delta_{\xi}$.
We introduce the  {\it symplectic current} $(n-1)$-form by anti-symmetrizing the variation of $\theta$ as 
\begin{align}
\Omega(\Phi,\delta \Phi,\delta_{\xi}\Phi)= \delta \theta (\delta_{\xi}\Phi)- \delta_{\xi} \theta (\delta \Phi).
\label{8}
\end{align}
Integrating $\Omega$ over the Cauchy surface gives 
\begin{align}
\delta H_{\xi}=\int_{\Sigma_t}\Omega(\Phi,\delta \Phi,\delta_{\xi}\Phi) = \int_{\Sigma_t}\delta J'_{\xi} 
\label{9}
\end{align} 
where the left-hand side defines the variation of the Hamiltonian $H_{\xi}$.
According to Eqs.(\ref{7}), (\ref{8}) and (\ref{9}) we have
\begin{align}
H_{\xi}=\int_{\Sigma_t}J'_{\xi}=\int_{\partial \Sigma_t}Q'[\xi],
\label{10}
\end{align}
using Stokes' theorem. The variation of $H_{\xi}$ should vanish on-shell. Considering one side of the Cauchy surface boundary to be on the bifurcation two-sphere and the other side to be at infinity,  Eq.(\ref{10}) gives 
\begin{align}
\int_{\mathcal{O}}\delta Q'[\xi]-\int_{\infty}\delta Q'[\xi]=0.
\label{11}
\end{align}

Now we want to obtain the BH entropy from this Noether charge. To do so, first we rewrite Eq.(\ref{6}) as
\begin{align}
J'_{\xi}= J_{\xi}-d(\xi\cdot l),
\label{12}
\end{align}
in which 
\be\label{12a}
J_{\xi}=\theta(\delta_{\xi}\Phi)-\xi\cdot L
\ee
 is the Noether current due to the Lagrangian $L$. Replacing $J'_{\xi}=dQ'[\xi]$ in Eq.(\ref{12}) gives
\begin{align}
J_{\xi}= d(Q'[\xi]+\xi\cdot l).
\label{13}
\end{align}
According to Eq.(\ref{13}) we have $J_{\xi}=dQ[\xi]$ in which $Q[\xi]=Q'[\xi]+\xi\cdot l$ and $dJ_{\xi}=0$. 

Consider $\xi^a=t^a+\mathit{\Omega}^{(\mu)}_H\phi^a_{(\mu)}$ as a Killing vector where the $\mathit{\Omega}^{(\mu)}_H$'s are the various angular velocities of the horizon of a multiply rotating black hole. Since the Killing vector vanishes on the bifurcation two-sphere, by inserting $Q'$ in terms of $Q$ in Eq.(\ref{11}) we have
\begin{align}
\int_{\mathcal O}\delta Q[\xi] = \int_{\infty} \delta Q[\xi]-\xi\cdot \delta l=\delta \mathcal{E} -\mathit{\Omega}^{(\mu)}_H\delta \mathcal{J}_{(\mu)},
\label{14}
\end{align}
in which the canonical energy $\mathcal{E}$ and angular momentum $\mathcal{J}$ can be defined as \cite{2}
\begin{align}
\mathcal{E}=\int_{\infty} Q[t]-t\cdot l,
\label{15}
\end{align}
and 
\begin{align}
\mathcal{J}_{(\mu)} =-\int_{\infty} Q[\phi_{(\mu)}]
\label{16}
\end{align}
and we see that $\mathcal{E}$, the ADM mass, depends on $l$, with the  integral  over the Cauchy surface boundary at infinity. We have assumed that   the space-time is asymptotically flat, and so  we can just consider Dirichlet BC;  $l$ will be unique for this choice on the Cauchy surface boundary.  Note that the term $\delta l.\phi$  does not appear in Eq.(\ref{15}) because $\phi^a$ is assumed to be tangent to the Cauchy surface boundary at infinity. In fact when $\xi^a$ is an asymptotic rotation,  we may choose the surface at infinity to be everywhere tangent to $\phi^a$, in which case the pullback of $\delta l \cdot \phi$ to that surface vanishes (on shell we have $\theta=\delta l$) \cite{3}.

Finally comparing Eq.(\ref{14}) with the second law of thermodynamics, we can obtain the BH entropy as 
\begin{align}
S=\frac{1}{T}\int_{\mathcal{O}} Q[\xi]=\frac{1}{T}\int_{\mathcal{O}} Q'[\xi],
\label{17}
\end{align}
where $T=\kappa/2\pi$ is the Unruh temperature and $\kappa=2\pi T$ is the surface gravity. 
 Note that the term containing $l$, i.e.the GHY term, is not involved in the BH entropy  because on the bifurcation two-sphere $\mathcal{O}$  the Killing vector $\xi^a=0$.  
We see that although a particular BC may change the GHY term, it does  not  modify the BH entropy. 
One can obtain Eq.(\ref{wald}) by using either the first or the second equality in Eq.(\ref{17}) \cite{7}.  

Note that although \eqref{15} depends on $l$, the first law of thermodynamics does not, since 
$T\delta S \int_{\mathcal O}\delta Q[\xi] = \int_{\infty} \delta Q[\xi]-\xi\cdot \delta l 
= \delta \mathcal{E} - \mathit{\Omega}^{(\mu)}_H\delta \mathcal{J}_{(\mu)}$.
 
\subsection{BH Entropy From Noether Charge in $f(R)$-gravity}

To illustrate these ideas, consider the BH entropy for $f(R)$-gravity.

We first begin by substituting $\mathcal{L}=f(R)/16\pi G$ in Eq.(\ref{wald}) yields 
\begin{equation}
\mathrm{S}=\frac{1}{4 G}\int_{\mathcal{O}} f'(R)\bar{\epsilon}
\label{18}
\end{equation}
where $f'(R)=\partial f(R)/\partial R$. For constant curvature $R_0$, it is well-known in $f(R)$-gravity that
\begin{equation}
R_0f'(R_0)-2f(R_0)=0\; .
\label{18'}
\end{equation} 
For a Schwarzschild BH $R_0=0$ implies $f(0)=0$  in\eqref{18'}, whereas  $f'(0)$ may have any constant value. Therefore the BH entropy reads
\begin{equation}
\mathrm{S}=\frac{f'(0)A}{4G},
\label{19}
\end{equation}
where $A$ is the surface of the Schwarzschild BH horizon \cite{12}.  In GR $f'(R)=1$, and so the BH entropy is $\mathrm{S}=A/4G$, which is the well-known Bekenstein-Hawking Formula \cite{8}. 

More generally, Schwarzschild- (anti) de Sitter solutions emerge in $f(R)$-gravity because of the action is non-linear in the Ricci scalar R  \cite{new1,new2}. From (\ref{18'}) we see that the solution $f(R_0)=\frac{f'(R_0)R_0}{2}$ where $R_0=4\Lambda$ is also permitted.  We shall consider this situation in appendix B. 
 
To further clarify the fact that the BH entropy is independent of BC, we shall now compute the BH entropy for $f(R)$-gravity   without using Eq.(\ref{wald}). First, we must make the Lagrangian of $f(R)$-gravity degenerate using Ostrogradsky approach. To do so we  introduce a scalar field $\phi$ and write the action of $f(R)$-gravity as\begin{equation}
\mathcal{A}_{f}=\int_\nu d^4x \sqrt{-g} f(R)=\int_\nu d^4x \sqrt{-g}(\phi R- V(\phi)),
\label{20}
\end{equation}
where $\phi=f'(R)$, $V(\phi)=R(\phi)\phi-f(R(\phi))$ and we have considered $f''(R)\neq 0$ \cite{18,19}. 

We know the Lagrangian of GR is degenerate through the following holographic relation \cite{20}
\begin{equation}
\sqrt{-g}R =\sqrt{-g}\mathcal{L}_{\textit{\tiny quad}}( g,\partial g)+\partial_i(\sqrt{-g}V^i),
\label{21}
\end{equation}
where
\[\mathcal{L}_{\textit{\tiny quad}}=\frac{1}{4}\mathcal{M}^{abcdef}
\partial_ag_{bc}\partial_dg_{ef},\]
\begin{equation}
\mathcal{M}^{abcdef}=g^{ad}(g^{bc}g^{ef}
-g^{be}g^{cf})+2g^{af}(g^{be}g^{cd}-g^{bc}g^{ed}),
\label{22}
\end{equation}
and 
\begin{equation}
\sqrt{-g}V^i=-g_{ab} \frac{\partial(\sqrt{-g}\mathcal{L}_{\textit{\tiny quad}})}{\partial(\partial_ig_{ab})}.
\label{23}
\end{equation}
Substituting Eq.(\ref{21}) in Lagrangian of Eq.(\ref{20}), we can find the degenerate Lagrangian of $f(R)$-gravity as \cite{15}
\begin{equation}
\sqrt{-g}f(R) =\sqrt{-g}\big(\phi\mathcal{L}_{\textit{\tiny quad}}+\partial_i\phi V^i-V(\phi)\big)+\partial_i(\sqrt{-g}\phi V^i).
\label{24}
\end{equation}
Taking  the variation of Eq.(\ref{24}) due to the infinitesimal coordinate transformation $x^a\rightarrow x^a+\xi^a $ gives \cite{21}
\begin{align}
\delta_{\xi}(\sqrt{-g}(\phi R- V(\phi))=&(E.O.M.)_{ab} \delta_{\xi} g^{ab}+(E.O.M.)_{\phi} \delta_{\xi}\phi+\nonumber \\
&\sqrt{-g}\nabla_a\bigg(\nabla_b\big(\phi(\nabla^b \xi^a-\nabla^a\xi^b)+2\xi^b\partial^a\phi-2\xi^a\partial^b\phi\big)\nonumber\\ 
&-2\xi^b \partial_b\partial^a\phi+2\xi^b \Box\phi+2\xi^b\phi g^{ad}R_{bd} \bigg),
\label{25}
\end{align}
where the equations of motions are
\begin{equation}
(E.O.M.)_{ab} \equiv \phi G_{ab}-\nabla_a\nabla_b\phi+\nabla_c\nabla^c\phi g_{ab}+\frac{V(\phi)}{2}g_{ab}=0
\end{equation}
and
\begin{equation}
(E.O.M.)_{\phi}\equiv \Box\phi-\frac{1}{3}\big(\phi \frac{dV(\phi)}{d\phi}-2V(\phi)\big)=0
\end{equation}
in which $G_{ab}$ is the Einstein tensor and we have used the relations, 
$\delta_{\xi}\phi=\xi^a \partial_a \phi$, $\delta_{\xi} g_{ab}=\nabla_a \xi_b+\nabla_b \xi_a$ and $\nabla_a \nabla_b \xi^i-\nabla_b \nabla_a \xi^i=R^i_{cab}\xi^c$. Comparing Eq.(\ref{25}) with Eq.(\ref{5}) we can obtain $\theta(\delta_{\xi}\Phi)$; substituting this into \eqref{12a} yields
\begin{equation}
J^a=\nabla_b\big(\phi(\nabla^b \xi^a-\nabla^a\xi^b)+2\xi^b\partial^a\phi-2\xi^a\partial^b\phi\big)\; .
\label{26}
\end{equation} 

From Eq.(\ref{26}) the Noether charge is
\begin{equation}
Q^{ab}=\phi(\nabla^b \xi^a-\nabla^a\xi^b)+2\xi^b\partial^a\phi-2\xi^a\partial^b\phi 
\label{27}
\end{equation}
which is anti-symmetric in the indices.  
Inserting Eq.(\ref{27}) in Eq.(\ref{17}) and retaining the coefficient $16\pi G$ we can 
\begin{equation}
S=\frac{1}{32\pi G T}\int_{\mathcal{O}}\bar{\epsilon} \hat{\epsilon_{ab}} \bigg( \phi(\nabla^b \xi^a-\nabla^a\xi^b)+2\xi^b\partial^a\phi-2\xi^a\partial^b\phi \bigg)=\frac{f'(0)A}{4G}
\label{28}
\end{equation}
for the  entropy for a Schwarzschild BH  in $f(R)$ gravity.
In obtaining the final result  (\ref{28}) we  made use of the fact that $\xi^a=0$, $\nabla^a\xi^b= \kappa \hat{\epsilon}^{ab}$ and $\hat{\epsilon_{ab}}\hat{\epsilon^{ab}}=-2$ on the bifurcation two-sphere $\mathcal{O}$,  and  that for a Schwarzschild BH with $R=0$, $\phi=f'(0)$ is constant. 

Note that had we considered $L'$ instead of $L$, then the entropy would not have  changed. Since $Q'[\xi]=Q[\xi]-\xi\cdot l$ and the Killing vector $\xi$ vanishes on $\mathcal{O}$, the entropy is unchanged by changing the BCs and the corresponding GHY terms,  though these may alter the Noether current and Noether charge.

Setting $\phi=1$ in the above calculations, we can easily extend our results to  GR.  Recall that for GR in four dimensions under Neumann BC,  one does not need to include any GHY terms to make the variational principle well-defined.  For $f(R)$-gravity this happens in $D$ dimensions under a special choice of mixed BCs \cite{15}. Therefore, for both cases, by substituting $l=0$ in Eqs.(\ref{11}) and (\ref{12}) we conclude for these BCs, in addition to the BH entropy, the Noether current and conserved charge would not change.
However for other boundary conditions $l\neq 0$, so the Noether current and conserved charge would change.  

\section{BH Entropy in Semi-Classical Approximation}

As outlined in section 2 throughout the Wald approach the BH entropy is independent of the BCs. 
Here we investigate whether or not the  the semi-classical approximation   exhibits this   as well. 

As in the Wald approach, in the semi-classical approximation it is important to have a well-defined variational principle \cite{6}. The partition function for an arbitrary gravitational model takes the form \cite{4}
\begin{equation}
\mathcal{Z}=\int [dg] e^{-\tilde{\mathcal{A}^*_E}} \simeq e^{-\tilde{\mathcal{A}^*_E}},
\label{29}
\end{equation}
where $\mathcal{A}^*_E=\mathcal{A}_E-\mathcal{A}_{E0}$ in which $\mathcal{A}_E $ is the Euclidean action and $\mathcal{A}_{E0} $ is the corresponding background Euclidean action. The tilde symbol denotes   dividing by $16\pi G$ and taking the limit $r \rightarrow \infty$, or in other words  
\begin{equation}
\tilde{\mathcal{A}^*_E}= \lim_{r\rightarrow \infty} \frac{\mathcal{A}^*_E}{16\pi G}.
\label{30}
\end{equation}
We can compute the free energy, energy and entropy 
\begin{align}
F&=\frac{-1}{\beta} ln \mathcal{Z}=\frac{1}{\beta}\tilde{\mathcal{A}^*_E},\nonumber \\
E&=F+\beta\frac{\partial F}{\partial \beta},\nonumber \\
S&=\beta^2 \frac{\partial F}{\partial \beta }
\label{31}
\end{align}
provided the action is well-defined.

As   in section 2, in the following we want to make the variational principle well-defined for $f(R)$-gravity and compute the entropy for different BCs. The key point in calculating the entropy for asymptotically flat space-times, is GHY terms \cite{22,23,24,25} -- these are different for different BCs and they depend on the space-time dimension \cite{15}. Yet we expect on physical grounds that   entropy should the same for all of them, since the black hole is a local object.
To solve this apparent inconsistency, let us continue with the $f(R)$-gravity action.

First we want to investigate the structure of $f(R)$-gravity action. Substituting Eq.(\ref{21}) into the action (\ref{20}) and taking variation gives \cite{15}
\begin{align}
\delta \mathcal{A}_{f}&= \int_\nu d^4 x \{ (E.O.M.)_{\phi}\delta \phi + (E.O.M.)^{ab} \delta g_{ab}\}  +\int_{-\Sigma_{t_1}+\Sigma_{t_2}} d^3y \bar{P}_\phi \delta \phi\nonumber \\
&+\frac{3}{2}\int_{-\Sigma_{t_1}+\Sigma_{t_2}} d^3y \sqrt{-g} \partial^0 \phi g^{ab}\delta g_{ab}+\frac{3}{2}\int_{\Gamma} d^3z \sqrt{-g} \partial^r \phi g^{ab}\delta g_{ab}\nonumber\\
& +\int_{\Gamma} d^3z \bar{\rho}_\phi \delta \phi-\int_{-\Sigma_{t_1}+\Sigma_{t_2}} d^3yg_{ab} \delta \bar{P}^{ab} -\int_{\Gamma} d^3z g_{ab} \delta \bar{\rho}^{ab} \nonumber \\
&  +3\int_{-\Sigma_{t_1}+\Sigma_{t_2}} d^3y \sqrt{-g} \delta(\partial^0 \phi)+ 3\int_{\Gamma} d^3z \sqrt{-g} \delta(\partial^1 \phi),
\label{32}
\end{align}
where the various surfaces are defined in (\ref{1}), and
\begin{align}
\bar{P}^{ab}&=\frac{\delta \mathcal{A}_{f}}{\delta(\partial_0 g_{ab})}= \phi P^{ab}+\frac{3}{2}\sqrt{-g} (g^{0i}g^{ab}-2g^{ib}g^{0a})\partial_i \phi, \nonumber \\
\bar{P}_\phi&=\frac{\delta \mathcal{A}_{f}}{\delta(\partial_0 \phi)}= g_{ab}P^{ab}=P, \nonumber \\
\bar{\rho}^{ab}&=\frac{\delta \mathcal{A}_{f}}{\delta(\partial_1 g_{ab})}= \phi \rho^{ab}+\frac{3}{2}\sqrt{-g} (g^{1i}g^{ab}-2g^{ib}g^{1a})\partial_i \phi, 
\nonumber \\
\bar{\rho}_\phi&=\frac{\delta \mathcal{A}_{f}}{\delta(\partial_1 \phi)}= g_{ab}\rho^{ab}=\rho.  
\label{33}
\end{align}
in which $\bar{P}^{ab}$, $\bar{P}_{\phi}$ are the canonical momenta of $g_{ab}$, $\phi$ in $f(R)$-gravity and $P^{ab}$ is the canonical momenta of $g_{ab}$ in GR \cite{15}. Note also
\[
g_{ab} P^{ab}= g_{ab}  \frac{\partial(\sqrt{-g} \mathcal{L}_{\textit{\tiny quad}})}{ \partial(\partial_0 g_{ab})}= \frac{1}{\sqrt{-g}} \partial_a (g g^{0a}),
\]
\begin{equation}
g_{ab} \rho^{ab}= g_{ab}  \frac{\partial(\sqrt{-g} \mathcal{L}_{\textit{\tiny quad}})}{ \partial(\partial_1 g_{ab})}=\frac{1}{\sqrt{-g}} \partial_a (g g^{1a}),
\label{34}
\end{equation}
 where $1$ denotes the component $r$ and here we have used the notation $\bar{P}^{ab}$, etc. to distinguish the quantities from the GR case. In what follows we will consider different kinds of BCs.
\subsection{Different Kind of Boundary conditions}
\subsubsection{Dirichlet BC}

 Under Dirichlet BC $\delta\phi\lvert_{\Sigma_{t_1},\Sigma_{t_2},\Gamma}=\delta g_{ab}\lvert_{\Sigma_{t_1},\Sigma_{t_2},\Gamma}=0$ and the four first terms of  (\ref{32}) vanish.  For asymptotically flat space-times $g_{ab}\to \eta_{ab}$ and $\phi\to $ constant as $r \rightarrow \infty $.   Eqs.(\ref{33}) and (\ref{34}) imply that we can omit the surface integral terms on the hyper-surface $\Gamma$ in  (\ref{32}). 
 
We therefore redefine the action by adding the following GHY terms 
\begin{align}
\mathcal{A}_{\textit{f(D)}}&=\mathcal{A}_{\textit{f}}+\mathcal{A}^{\textit{GHY}}_{\textit{f(D)}}=\int_\nu d^4x \sqrt{-g} f(R)+\int_{-\Sigma_{t_1}+\Sigma_{t_2}} d^3yg_{ab} \bar{P}^{ab}\nonumber \\
&-3\int_{-\Sigma_{t_1}+\Sigma_{t_2}} d^3y \sqrt{-g} \partial^0 \phi
\label{35}
\end{align}
so that $\mathcal{A}_{\textit{f(D)}}$  is well-defined at $r \rightarrow \infty$. The GHY terms in Eq.(\ref{35}) are the well-known GHY terms in the ADM method, since inserting the expression \cite{15}
\begin{equation}
g_{ab} P^{ab}= \sqrt{-g}[-2Kn^0+\frac{\partial _\alpha N^\alpha}{N}],
\label{36}
\end{equation}
in the surface terms of Eq.(\ref{35}) and substituting $\phi=f'(R)$ leads to 
\begin{align}
\mathcal{A}^{\textit{GHY}}_{\textit{f(D)}}&=\int_{-\Sigma_{t_1}+\Sigma_{t_2}} d^3y \sqrt{h} f'(R)K+\int_{-\Sigma_{t_1}+\Sigma_{t_2}} d^3y f'(R) \frac{\partial _\alpha N^\alpha}{N},
\label{37}
\end{align}
where $h$ and $\mathit{K}$ are the trace of the induced metric and the extrinsic curvature; $n^0$ denotes the component $t$ of normal vector  $n$ to $\Sigma_{t}$ and the lapse and shift functions are denoted by $N$ and $N^{\mu}$ respectively. Note that terms containing variations of $N$ and $N^\alpha$ on the $\Sigma_{t_1}$ and $\Sigma_{t_2}$ vanish under Dirichlet (i.e. $\delta h^{ab}|_{\tiny{\text{Boundary}}}=\delta N^{\mu}|_{\tiny{\text{Boundary}}}=\delta N|_{\tiny{\text{Boundary}}}=0 $).

We now turn back to the relation (\ref{35}) to calculate the entropy for the $f(R)$-gravity action. For the special case of a Schwarzchild BH, we  compare the  value of the Euclidean action for the perturbed system to that of the flat background space-time.  Replacing $t\rightarrow i\tau$ in the Schwarzschild 
metric we have
\begin{equation}
dS_E^2= (1-\frac{2GM}{r})d\tau^2+(1-\frac{2GM}{r})^{-1}dr^2+r^2d\Omega^2,
\label{38}
\end{equation}
in which the parameter $\tau$ is periodic as $\tau=\tau+\beta ;\beta=8\pi GM$. The period of $ \tau $ is obtained from the Unruh temperature formula $\beta = 2\pi/\kappa$ and for the Schwarzschild metric $\kappa=\frac{1}{4GM}$ \cite{13}. The  value of the Euclidean action (\ref{35}) vanishes for both the Schwarzschild metric and   the background flat metric\footnote{The bulk contribution for both metrics vanishes, because for $R=0$ as we reveal in Eq.(\ref{19}) we have $f(0)=0$ and $f'(0)=\text{constant}$. Also since $\phi=f'(0)$, we have $\partial^0\phi=\partial^1\phi=0$. Therefore, for both metrics the surface terms on $\Sigma_{t_1}$ and $\Sigma_{t_2}$ vanish following  (\ref{33}) and (\ref{34}).}. 

So according to the relation (\ref{30}),  $\tilde{A}^*_E$ is zero.  The origin of this strange result comes from inappropriately discarding the surface integral terms associated with $\Gamma$ in (\ref{32}),
implying that we are not allowed to use (\ref{35}) to obtain  the BH entropy.  To obtain the value of the Euclidean action for Schwarzschild BH in the region $2G M<r<\infty$, we must render the action well-defined in the region 
of  large but finite $r$.  In this case the surface integral terms associated with $\Gamma$ in Eq.(\ref{32}) 
do not vanish, and we consider the action
\begin{align}
\mathcal{A}_{\textit{f(R)(D)}}&=\mathcal{A}_{\textit{f(R)}}+\mathcal{A}^{\textit{GHY}}_{\textit{f(R)(D)}}=\int_\nu d^4x \sqrt{-g} f(R)+\int_{-\Sigma_{t_1}+\Sigma_{t_2}} d^3yg_{ab} \bar{P}^{ab}\nonumber \\
&-3\int_{-\Sigma_{t_1}+\Sigma_{t_2}} d^3y \sqrt{-g} \partial^0 \phi+\int_{\Gamma} d^3z g_{ab} \bar{\rho}^{ab}-3\int_{\Gamma} d^3z \sqrt{-g} \partial^1 \phi.
\label{39}
\end{align}
Similar to Eq.(\ref{36}) we can show
\begin{equation}
g_{ab} \rho^{ab}=\sqrt{-g}[-2\mathcal{K}r^1+\frac{\partial _\alpha N^{\alpha}}{N}],
\label{40}
\end{equation}
in which $r^1$ denotes the component $r$ of normal vector $r^n$ to $\Gamma$. 
Hence, in the framework of ADM formulation,  using Eqs.(\ref{33}), (\ref{36}) and (\ref{40}), we can rewrite the GHY terms in the Eq.(\ref{39}) as follows
 \begin{align}
\mathcal{A}^{\textit{GHY}}_{\textit{f(R)(D)}}&=\int_{-\Sigma_{t_1}+\Sigma_{t_2}} d^3y \sqrt{h} f'(R)K+\int_{\Gamma} d^3z \sqrt{\gamma} f'(R)\mathcal{K},
\label{41}
\end{align}
where $ \gamma$ is the determinant of the induced metric and $\mathcal{K}$ is the extrinsic curvature on $\Gamma$. In Eq.(\ref{41}) it is assumed that we are in a coordinate frame where $N^\alpha=0$ and $N=1$ on the boundary $\Gamma$; this assumption is not necessary but simplifies the equations. This GHY term makes the action well-defined either at finite $r$ on $\Gamma$ or for $r \rightarrow \infty$. 

In order to calculate BH entropy we consider the Schwarzschild BH as a solution for $f(R)$-gravity and the flat metric as the background. Since $R=0$ for both cases, the bulk terms in Eq.(\ref{39}) vanish. Also remember $\phi=f'(0)=\text{constant}$; so we have $\partial^0\phi=\partial^1\phi=0$. Therefore, the contribution of surface terms on $-\Sigma_{t_1}+\Sigma_{t_2}$ and the last term in Eq.(\ref{39}) will vanish. Hence, the only term which gives entropy is the following
\begin{equation}
\int_\Gamma d^3z \phi g_{ab}\rho^{ab},
\label{42}
\end{equation}
where using Eq.(\ref{33}), $\bar{\rho}^{ab}$ is changed to $\phi \rho^{ab} $ due to constancy of $\phi$. Considering $\phi=f'(0)$, after some algebra (see Appendix A) we have, using (\ref{38})  
\begin{equation}
\tilde{\mathcal{A}}^*_E=\lim_{r\rightarrow \infty}\frac{\mathcal{A}^*_E}{16\pi G}=\frac{f'(0)\beta^2}{16\pi G},
\label{43}
\end{equation}
which gives finally  \begin{equation}
F=\frac{\beta f'(0)}{16\pi G},\hspace{1cm}E=\frac{f'(0)\beta}{8 \pi G}=f'(0)M,\hspace{1cm}S=\frac{\beta^2}{16\pi G}=\frac{f'(0)A}{4G},
\label{44}
\end{equation}
using \eqref{31} and the expression for $\beta$, where $A=4\pi(2 G M)^2$.  
This coincides with the known result of Wald method. 

Our results to the GR case by replacing $\phi=1$ and $V(\phi)=0$ in all calculations. Then the Schwarzschild BH entropy $S=A/4G$ is obtained for  Dirichlet BC.


\subsubsection{Neumann BC}

Consider the Neumann BC $\delta \bar{P}^{ab}\lvert_{\Sigma_{t_1},\Sigma_{t_2}}=\delta \bar{P}_\phi\lvert_{\Sigma_{t_1},\Sigma_{t_2}}=0$. Moreover, for asymptotically flat space-time on the boundary $\Gamma$ we need only consider the Dirichlet BC $\delta g_{ab}\lvert_\Gamma=\delta\phi\lvert_\Gamma=0$. Fortunately this is consistent\footnote{We know the hyper-surfaces $\Sigma_{t}$ intersect $\Gamma$ orthogonally. Thus $r_a n^a=0$ and the BCs on $\Sigma_{t}$ and $\Gamma$ are independent.} with the Neumann BCs. Hence, using the key Eq.(\ref{32}), we should add the following GHY term to the action \cite{15}
\begin{align}
\mathcal{A}_{\textit{f(N)}}&=\mathcal{A}_{\textit{f}}+\mathcal{A}^{\textit{GHY}}_{\textit{f(N)}}=\int_\nu d^4x \sqrt{-g} f(R)-\int_{-\Sigma_{t_1}+\Sigma_{t_2}} d^3y \bar{P}_\phi \phi \nonumber \\
&-3\int_{-\Sigma_{t_1}+\Sigma_{t_2}} d^3y \sqrt{-g} \partial^0 \phi+\int_{\Gamma} d^3z g_{ab} \bar{\rho}^{ab}-3\int_{\Gamma} d^3z \sqrt{-g} \partial^1 \phi.
\end{align}
In comparing the Schwarzschild BH with the flat background metric, similar reasons as given in Dirichlet case indicate that the bulk term, the integrals on $-\Sigma_{t_1}+ \Sigma_{t_2}$ and the terms containing  $\partial^0\phi$ and $\partial^1\phi$ do not contribute to the entropy. The remaining term of $\mathcal{A}_{\textit{f(N)}}$ is once again $\int_\Gamma d^3z \phi g_{ab}\rho^{ab}$, using \eqref{33},
and this yields the same entropy as obtained for Dirichlet BC in Eq.(\ref{44}). 

\subsubsection{Mixed BC}

It has been shown \cite{15} that   $f(R)$-gravity may also be described consistently by two distinct types of BCs\\

\noindent
i) $\delta \bar{P}^{ab}\lvert_{\Sigma_{t_1},\Sigma_{t_2}}=\delta \phi \lvert_{\Sigma_{t_1},\Sigma_{t_2}}=0$, $\delta \phi \lvert_\Gamma=\delta g_{ab}\lvert_{\Gamma}=0$\\
ii) $\delta \bar{P}_\phi\lvert_{\Sigma_{t_1},\Sigma_{t_2}}=\delta g_{ab}\lvert_{\Sigma_{t_1},\Sigma_{t_2}}=0$ ,    $\delta \phi \lvert_\Gamma=\delta g_{ab}\lvert_{\Gamma}=0$\\

Each kind of the above BCs requires its own GHY terms. Hence, the action corresponding to the above BCs read respectively as
\begin{align}
\mathcal{A}_{\textit{f(M1)}}&=\mathcal{A}_{\textit{f}}+\mathcal{A}^{\textit{GHY}}_{\textit{f(M1)}}=\int_\nu d^4x \sqrt{-g} f(R)-3\int_{-\Sigma_{t_1}+\Sigma_{t_2}} d^3y \sqrt{-g} \partial^0 \phi \nonumber\\
&+\int_{\Gamma} d^3z g_{ab} \bar{\rho}^{ab}-3\int_{\Gamma} d^3z \sqrt{-g} \partial^1 \phi,
\end{align}
\begin{align}
\mathcal{A}_{\textit{f(M2)}}&=\mathcal{A}_{\textit{f(R)}}+\mathcal{A}^{\textit{GHY}}_{\textit{f(M2)}}=\int_\nu d^4x \sqrt{-g} f(R)+\int_{\Gamma} d^3z g_{ab} \bar{\rho}^{ab}\nonumber\\
&-3\int_{\Gamma} d^3z \sqrt{-g} \partial^1 \phi.
\end{align}
Due to similar reasons as mentioned for the Dirichlet and Neumann cases none of the terms make any contribution to the BH entropy except for the term given in Eq.(\ref{42}). So the thermodynamic quantities are be the same as derived in Eq.(\ref{44}). 

As is observed, for asymptotically flat solutions of $f(R)$-gravity, regardless of the kind of BCs, i.e. Dirichlet, Neumann or mixed, the term responsible to give the numerical value of the action is the same. Therefore, the physical quantities corresponding to a BH solution would be the same.
 
Note that in order to obtain the equations of motion via the second mixed BC, we need not add  GHY terms to the action for an asymptotically flat space-time in arbitrary dimension \cite{15}; indeed
the equations of motion are insensitive to the choice of boundary terms. One might think that since the terms on the lateral boundary $\Gamma$ vanish at special infinity $r\rightarrow\infty$ (for an asymptotically flat space-time) that they can be discarded, but in fact this is not the case.  These terms at finite $r$ are essential for 
computing the BH entropy, and should be incorporated in the $f(R)$-gravity action.  Including such terms renders the action well-defined at finite $r$.


For GR, i.e. $\phi=1$ and $V(\phi)=0$ the first mixed BC reduces to the Neumann BC $\delta P^{ab}\lvert_{\Sigma_{t_1},\Sigma_{t_2}}=0$, $\delta g_{ab}\lvert_{\Gamma}=0$, Hence, the appropriate GHY term is 
\begin{align}\label{353}
\mathcal{A}^{\textit{GHY}}_{\textit{GR(N)}}=\int_{\Gamma} d^3z g_{ab} \rho^{ab}.
\end{align}
As noticed above to attain the equations of motion for asymptotically flat space-time\footnote{For asymptotically flat space-time we have $g_{ab}=\eta_{ab}$ at the limit $r \rightarrow \infty $. Hence, using Eqs.(\ref{33}) and (\ref{34}) by considering $\phi=1$ for GR, the surface integral terms on $\Gamma$ disappear.} we need not to add this GHY term to the action \cite{15,16}, however, for calculation BH entropy adding this term is essential. Then by repeating similar calculations as before we find the same result for $\tilde{\mathcal{A}^*_E}$ under Dirichlet BC in GR. In this way, the partition function, Helmholtz free energy, energy and entropy have the same value regardless of the kind of BC, i.e. Neumann or Dirichlet in GR.

 \section{Conclusion}

In this paper, we computed the BH entropy, in the framework of Wald method and Euclidean semi-classical approximation via different BCs in higher curvature gravity such as $f(R)$-gravity and special case GR \cite{1,2,3,4,5,6}. In the Wald method, at first we considered a well-defined gravitational action under arbitrary BCs. We showed that although the definition of Noether current and Noether charge can alter under different BCs through the appropriate GHY terms,  the entropy of a black hole does not change. 
Consequently the Wald formula can be employed to compute BH entropy from the Lagrangian density under different BCs  \cite{6,7}. As an example we demonstrated this for $f(R)$-gravity and GR.

In Euclidean semi-classical approximation the BH entropy may be obtained for different kinds of BCs and GHY terms. However, we showed for asymptotically flat BH solutions such as Schwarzschild, the main term  that is responsible for giving the difference of the  value of the action relative to the background solution  is the same for all cases, regardless of the particular kind of BCs. In fact, decomposing the space-time boundary into one time-like $\Gamma$ and  two space-like $\Sigma_t$ hyper-surfaces, one can see that the integral over $\Gamma$ has no role in making the action principle well-defined for asymptotically flat metrics \cite{15} when the boundary is taken to be spatial infinity.  However if we require that the variational principle be well-posed quasi-locally (at finite $r$) then we should retain the boundary terms on the lateral boundary $\Gamma$.  This yields the BH entropy, and the integral over $\Gamma$ is exactly the term that gives non trivial contributions for different BCs. Taking this point into account, we showed that in $f(R)$-gravity, under Dirichlet, Neumann and two types of mixed BCs and also in GR, under Dirichlet and Neumann BCs, the entropy does not change and is the same as we obtained in the Wald method. 

It is straightforward to extend our considerations to   asymptotically  AdS  space-times in $f(R)$-gravity (see Appendix B). The interesting point is that in this case the bulk term yields the BH entropy and the GHY terms make no contribution. In considering an AdS metric as the background, needs to change the ensemble  from canonical (for the case of flat background) to isothermal-isobaric.
   
For simplicity and clarity, we considered the problem for the Schwarzschild metric in the flat background in $D=4$. Our method is straightforwardly generalizable to arbitrary dimensions and different types of asymptotically flat BH solutions.  For  example, in the semi-classical approximation in each case we just need to compute 
$$
-2\int_\Gamma d^{D-1}z\sqrt{\gamma}(\mathcal{K}-\mathcal{K}_0)
$$  for GR and 
$$-2\int_{\Gamma} d^{D-1}z\sqrt{\gamma}f'(R)(\mathcal{K}-\mathcal{K}_0)
$$
for $f(R)$-gravity in D dimensions.

\section*{Acknowledgments}

 The authors would like to thank the Iran National Science Foundation (INSF) for supporting this research under grant number 97015575. H.Khodabakhshi thanks Ali Naseh for useful discussions. F.Shojai is grateful to the University of Tehran for supporting this work under a grant provided by the university research council.

\appendix
\numberwithin{equation}{section}
\section{Numerical Values of Euclidean Action}

As we mentioned the  term that gives us BH entropy in $f(R)$-gravity is  Eq.(\ref{42}). Using Eq.(\ref{40}) we can express this term in terms of ADM variable and also we can assume a coordinate frame where $N^\alpha=0$ and $N=1$ on the boundary $\Gamma$. Since for background metric and Schwarzschild BH $R=0$ and substituting $\phi=f'(0)=\text{constant}$, finally we have
\begin{equation}
\mathcal{A}^*_E=\mathcal{A}_E-\mathcal{A}_{E_0}=-2f'(0)\big(\int_{\Gamma}d^{3}z\sqrt{\gamma}\mathcal{K}-\int_{\Gamma}d^{3}z\sqrt{\gamma_0}\mathcal{K}_0\big).
\label{a1}
\end{equation}
By means of Eq.(\ref{38}), the reduced metric for BH and in $\gamma_{ab}$ background is as follows
\begin{equation}
dS_{\Gamma}^2=(1-\frac{2GM}{r})d\tau^2+r^2d\Omega^2,
\label{a2}
\end{equation}
and 
\begin{equation}
dS^2_{\Gamma_0}=d\tau^2+r^2d\Omega^2,
\end{equation}
where their rotational parts are the same. In order to make the boundary unified for two metrics, we should match length of the circle of Euclidean time, i.e. \cite{23}
\begin{equation}
\int_0^\beta d\tau\sqrt{\gamma}=\int_0^{\beta_0}d\tau\sqrt{\gamma_0}.
\label{a3}
\end{equation}
Then $\beta_0=\sqrt{1-2GM/r}\beta$. Also $\sqrt{\gamma}=r^2\sin \theta\sqrt{1-2GM/r}$ and $\sqrt{\gamma_0}=r^2\sin\theta$ and we have
\begin{equation}
\mathcal{K}=\mathcal{K}^{ab}\gamma _{ab}=\frac{GM}{r^2}\frac{1}{\sqrt{1-2GM/r}}+\frac{2}{r}\sqrt{1-2GM/r},
\label{a4}
\end{equation}
and 
\begin{equation}
\mathcal{K}_0=\frac{2}{r},
\end{equation}
and $d^3z=d \tau d \theta d \phi$. Then we have
\begin{equation}
\int_\Gamma d^3z \sqrt{\gamma} \mathcal{K}=\int_0 ^{8\pi GM} d\tau \int_0 ^{2\pi} d \phi \int_0 ^\pi d\theta \sin \theta (2r-3GM)=4\pi \beta (2r-3GM).
\label{a5}
\end{equation} 
In order to compute the term consisting $\mathcal{K}_0$ for the asymptotic flat space-time
\begin{equation}
\int_\Gamma d^3z \sqrt{\gamma_0} \mathcal{K}_0= 8\pi \beta r \sqrt{(1-2GM/r)}.
\label{a6}
\end{equation}
Adding up our results in Eqs.(\ref{a1}), (\ref{a5}) and (\ref{a6}) and letting $r\rightarrow \infty$ and inserting the factor $1/16\pi G$ we have 
\begin{equation}
\tilde{\mathcal{A}}^*_E=\lim_{r\rightarrow \infty}\frac{\mathcal{A}^*_E}{16\pi G}=-\frac{f'(0)}{8\pi G}\lim_{r\rightarrow \infty}\int_\Gamma d^3z \sqrt{\gamma}(\mathcal{K}-\mathcal{K}_0)=\frac{f'(0)\beta^2}{16\pi G}.
\end{equation}
using (\ref{38}).

\section{Asymptotically AdS space-times}

We want to show for an asymptotic AdS space-time, unlike asymptotic flat spaces, the term that gives the entropy is not the GHY term, but is a bulk term. To see  this point, consider the Euclidean Schwarzschild AdS solution of $f(R)$-gravity as
\begin{equation}
dS^2_E=(1-\frac{2GM}{r}+\frac{r^2}{b^2})d\tau^2+(1-\frac{2GM}{r}+\frac{r^2}{b^2})^{-1}dr^2+r^2d\Omega^2,
\label{A1}
\end{equation}
where $\Lambda=-3/b^2$ and $\tau=\tau+\beta$ while $\beta=4\pi b^2r_+/(b^2+3r_+^2)$ in which $r_+$ is radius the event horizon defined as $2GM=r_+(r_+^2+b^2)/b^2$. 

The background metric is achieved by assuming $M=0$, which is the ordinary AdS metric as 
\begin{equation}
dS^2_{E_0}=(1+\frac{r^2}{b^2})d\tau^2+(1+\frac{r^2}{b^2})^{-1}dr^2+r^2d\Omega^2,
\label{A2}
\end{equation}
where $\tau=\tau+\beta_0$ and from the matching condition
\begin{equation}
\int_0^\beta d\tau\sqrt{\gamma}=\int_0^{\beta_0}d\tau\sqrt{\gamma_0},
\label{A3}
\end{equation}
the   periods are related via 
\begin{equation}
\beta_0=\sqrt{1-\frac{2Mb^2}{b^2r+r^3}}\beta.
\end{equation}

Consider an AdS Schwarzschild BH and an AdS background, both having the same curvature scale $R_0$. It might seem that the bulk terms cancel as usual and the entropy comes out from the GHY term, but this  is not the case. Let us first calculate the contribution of the GHY term for this model. We learned  in the previous discussion that the regardless of the choice of BC, the following expression contributes to the quantity $\mathcal{A}^*_E$ as
\begin{equation}
\mathcal{A}^*_E=\mathcal{A}_E-\mathcal{A}_{E_0}=-2f'(R_0)(\int_{\Gamma} d^3 z\sqrt{\gamma}\mathcal{K}-\int_{\Gamma} d^3z\sqrt{\gamma_0}\mathcal{K}_0),
\label{A4}
\end{equation}
where we have used the fact that $f'(R_0)$ is constant. Using Eqs.(\ref{A1}) and (\ref{A2}) we have
at fixed large $r$
\begin{align}
\mathcal{A}^*_E=&-2f'(R_0)\left\{ \int_0^\beta d\tau\int_0^{2\pi}d\phi\int_0^\pi\sin\theta d\theta(2r-3GM+\frac{3r^3}{b^2}) \right. \nonumber\\
-& \left. \int_0^{\beta_0}d\tau\int_0^{2\pi}d\phi\int_0^{\pi}\sin \theta d\theta(2r+\frac{3r^3}{b^2})\right\}\nonumber\\
=&-8\pi\beta f'(R_0)(2r-3GM+\frac{3r^3}{b^2}-\sqrt{1-\frac{2Mb^2}{b^2 r+r^3}}(2r+\frac{3r^3}{b^2})).
\label{A5}
\end{align}
Taking the limit $r\rightarrow r_{\infty}$ we find
\begin{equation}
\tilde{\mathcal{A}^*}_{\tiny{E (GHY)}}=\lim_{r\rightarrow r_{\infty}} \frac{  \mathcal{A}^*_E}{16\pi G}=0.
\label{A6}
\end{equation}
Hence, unlike asymptotically flat spaces, the GHY term does not contribute in BH entropy.  Hence the well-posed action of $f(R)$-gravity reduces to a volume integral $\int_{ \nu } d^4x\sqrt{-g} f(R)$. The important point is although the integrand of the bulk term is the same for the BH and background, the measures are not the same. This is because the periods $\beta$ and $\beta_0$ differ and integration range on  $r$ also differs if it is up to
some  cutoff value at fixed large $r$.   We obtain
\begin{align}
\mathcal{A}^*_E &=  \mathcal{A}_E- \mathcal{A}_{E_0}=f(R_0)\{\int_0^\beta d\tau\int_{r_+}^{r}r^2dr\int_0^{2\pi}\sin\theta d\theta\nonumber\\
&-\int_0^{\beta_0}d\tau\int_0^{r}r^2dr\int_0^{2\pi}\sin \theta d\theta\}\nonumber\\
&=\frac{4\pi\beta}{3}f(R_0)(r^3-r^3_+-\sqrt{1-\frac{2GMb^2}{b^2r+r^3}}r^3).
\label{A7}
\end{align}
Using equation of motion, we have $f(R_0)=\frac{f'(R_0)R_0}{2}$ where $R_0=4\Lambda$. So 
\begin{equation}
\tilde{\mathcal{A}}^*_E=  {\frac{f'(R_0) \beta}{2b^2G}}(r_+^3- GMb^2).
\label{A8}
\end{equation}
upon dividing by $16\pi G$.

We can express the free energy in terms of $r_+$ is 
\begin{align}
F=\frac{1}{\beta}\tilde{\mathcal{A}}_E^*&=\frac{f'(R_0)}{4b^2G}(r_+^3-br_+).
\label{A9}
\end{align}
This result is different from that of free energy in GR and $f(R)$-gravity. However, note that the free energy may be either Helmholtz or Gibbs free energy. In the above considerations we inserted the cosmological constant by considering the AdS as the background metric. As is well-known the cosmological constant introduces a constant pressure as  {\cite{29}}  
\begin{equation}
P=-\frac{\Lambda}{8\pi G}.
\end{equation}
When we consider a flat  background  metric  at constant temperature and volume, the partition function would give us the Helmholtz free energy $F(T,V)$ and we are in the canonical ensemble. However,
in our calculations concerning the AdS Schwarzschild BH we have somehow constant  temperature and pressure, instead. This conditions resembles the isothermal-isobaric conditions and the partition function would give us the Gibbs free energy $\mathcal{G}(T,P)$ \cite{30}.  Hence, the free energy $F$ in the Eq.(\ref{A9}) is the Gibbs free energy and the entropy comes out as 
\begin{equation}
\mathit{S}=\beta^2\frac{\partial \mathcal{G}}{\partial \beta}=\frac{f'(R_0)A}{4G},
\label{A10}
\end{equation}
where $A=4\pi r_+^2$ is the area of the horizon. Also using the Legendre transformation  $\mathcal{G}=H-TS$ the total enthalpy reads
\begin{equation}
H=\mathcal{G}+TS=f'(R_0)M.
\end{equation}
As we see the type of thermodynamical ensemble  may differ for different metrics, i.e. canonical ensemble for the flat background and isothermal-isobaric ensemble for AdS background.

\end{document}